\documentclass[10pt,twocolumn,twoside]{IEEEtran}
\usepackage{amsmath,amsfonts}
\usepackage{algorithmic}
\usepackage{algorithm}
\usepackage{array}
\usepackage[caption=false,font=normalsize,labelfont=sf,textfont=sf]{subfig}
\usepackage{textcomp}
\usepackage{stfloats}
\usepackage{url}
\usepackage{verbatim}
\usepackage{graphicx}
\usepackage{cite}
\usepackage{amssymb}
\usepackage{soul} 
\usepackage{color, xcolor} 
\usepackage{multirow}
\usepackage{CJKutf8}

\begin{document}

\title{AI-Driven Mobility Management for 
High-Speed Railway Communications: Compressed Measurements and Proactive Handover}

\author{Wen Li, Wei Chen,~\IEEEmembership{Senior Member,~IEEE,} Shiyue Wang, Yuanyuan Zhang, \\Michail Matthaiou,~\IEEEmembership{Fellow,~IEEE}, and Bo Ai,~\IEEEmembership{Fellow,~IEEE}

\thanks{
Wen Li, Wei Chen, Shiyue Wang and Bo Ai are with the State Key Laboratory of Advanced Rail Autonomous Operation, Beijing Jiaotong University, China and the School of Electronic and Information Engineering, Beijing Jiaotong University, Beijing, China. (e-mail: 24120062@bjtu.edu.cn; weich@bjtu.edu.cn; shiyuewang@bjtu.edu.cn; boai@bjtu.edu.cn). Corresponding author: Wei Chen

Y. Zhang is with the Mediatek (Beijing) Inc, Beijing, China. (e-mail: yuany.zhang@mediatek.com)

M. Matthaiou is with the Centre for Wireless Innovation (CWI), Queen’s University Belfast, UK. (e-mail: m.matthaiou@qub.ac.uk)

}
}



\maketitle

\begin{abstract}
High-speed railway (HSR) communications are pivotal for ensuring rail safety, operations, maintenance, and delivering passenger information services. The high speed of trains creates rapidly time-varying wireless channels, increases the signaling overhead, and reduces the system throughput, making it difficult to meet the growing and stringent needs of HSR applications. In this article, we explore artificial intelligence (AI)-based beam-level and cell-level mobility management suitable for HSR communications. Particularly, we propose a compressed spatial multi-beam measurements scheme via compressive sensing for beam-level mobility management in HSR communications. In comparison to traditional down-sampling spatial beam measurements, this method leads to improved spatial-temporal beam prediction accuracy with the same measurement overhead. Moreover, we propose a novel AI-based proactive handover scheme to predict handover events and reduce radio link failure (RLF) rates in HSR communications. Compared with the traditional event A3-based handover mechanism, the proposed approach significantly reduces the RLF rates which saves 50\% beam measurement overhead.

\end{abstract}

\begin{IEEEkeywords}
Artificial intelligence, compressive sensing, mobility management.
\end{IEEEkeywords}

\section{Introduction}
\IEEEPARstart{H}{igh}-speed railway (HSR) provides people with safe, fast, comfortable and economical modes of transportation, and has been widely developed around the world. Nevertheless, mobile communication systems are pivotal for ensuring rail safety, operations, maintenance, and delivering passenger information services. However, the high speed of trains entails formidable challenges due to the time-varying nature of wireless channels in HSR communications. These challenges include the increased signaling overhead and reduced system throughput.

\begin{figure}[!t]
\centering
\includegraphics[width=3.2in]{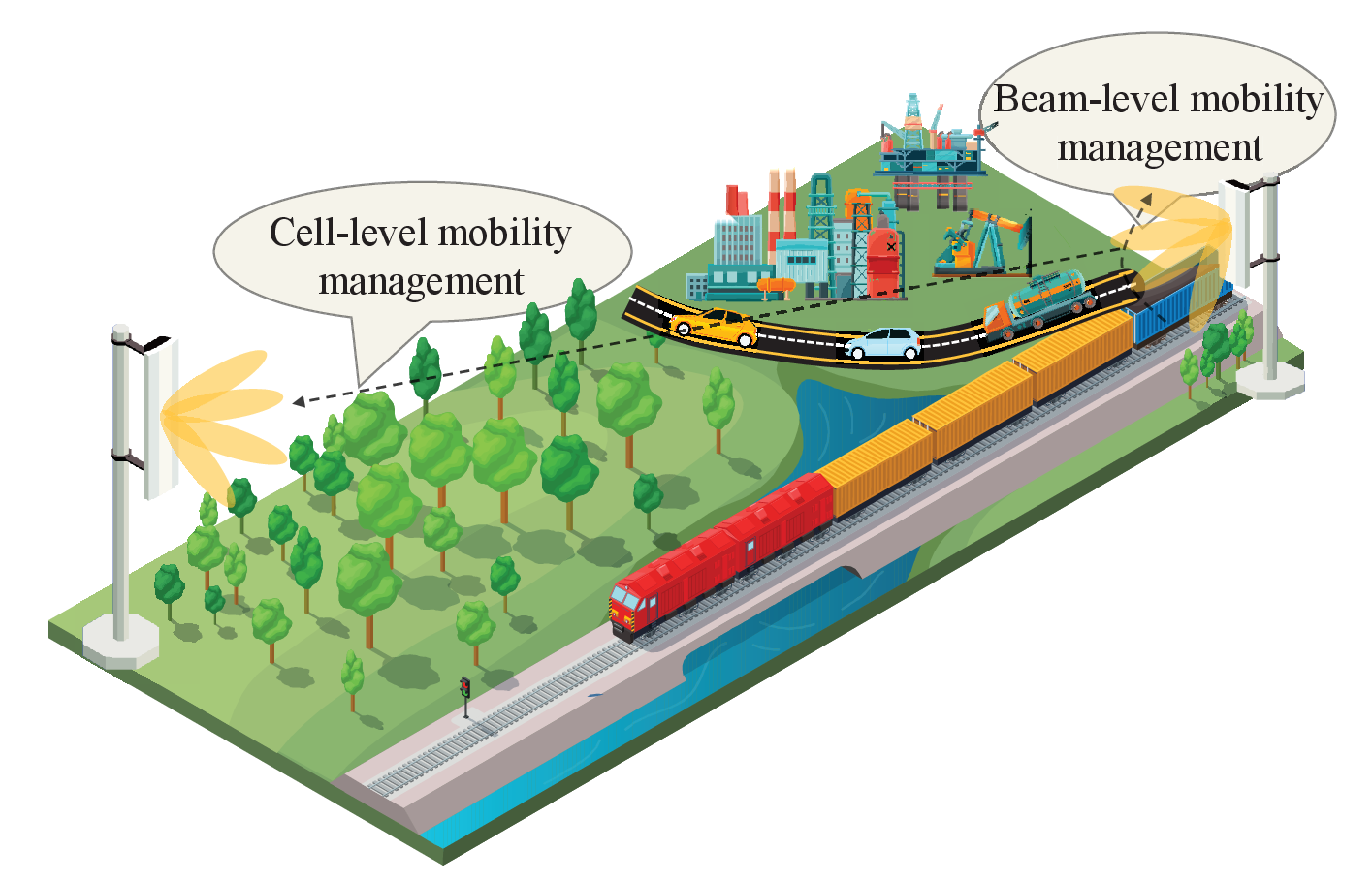}
\caption{Illustrations of beam-level mobility management and cell-level mobility management in HSR communications.}
\label{scenario diagramig}
\end{figure}

Thanks to the abundant spectrum available at higher frequencies, communication systems using the millimeter wave (mmWave) technology avail of faster data transmission speeds and reduced latency compared to traditional low-frequency communications. For HSR communications, mmWave frequency bands can underpin increased data transmission rates that meet the demands of onboard applications and services, and provides low latency that is indispensable for real-time communication requirements. However, mmWave communication systems suffer from significant path loss, which restricts the effective range of transmission. To tackle this problem, beamforming with large-scale antenna arrays can be employed to generate finely directional beams, mitigating propagation loss and expanding signal coverage. However, the high mobility of trains entails pressing challenges in mobility management for HSR communications. While being connected, a train has to address two types of mobility, i.e., the beam-level and cell-level mobility as specified in the 3rd Generation Partnership Project (3GPP) Release 17 for fifth generation (5G) systems\cite{3gpp2020}. Beam-level mobility involves dynamically switching beams during which the train establishes a link path to a new beam within the same cell, ensuring communication with high performance. On the other hand, cell-level mobility focuses on cell handover between different cells. As trains move from one coverage zone to another, it is crucial to have a smooth cell handover to sustain a continuous connectivity experience (Figure \ref{scenario diagramig}).

Beam-level mobility management in 5G relies on periodic channel state information reference signals and synchronization signal blocks for beam sweeping, measurement, and reporting. However, frequent beam sweeping introduces significant latency and high overhead\cite{10123939}, making it impractical for HSR communications. Cell-level mobility management primarily relies on an event A3-based handover process, encompassing handover trigger, preparation, and execution. Before handover completion, the UE continues to communicate with the source base station (BS), often resulting in a sharp deterioration in the communication quality. Additionally, with a fixed handover period, the link quality between the high-speed terminal and source BS deteriorates faster, increasing the risk of handover failure. Hence, innovative handover mechanisms are urgently needed to address these limitations.

In recent years, deep neural networks (DNNs) have proven their effectiveness in capturing the complex and nonlinear characteristics of wireless channels to enhance mobility management. For example, in \cite{10123939} the optimal beam for future instances was predicted by using a multi-layer long short term memory (LSTM) network. The appropriate BS for cell handover was predicted by using convolutional neural network (CNN)-based auto-encoders and LSTM networks in \cite{shah2022multi}. However, these methods either measure all beams, resulting in high overhead, or measure a part of all beams, resulting in notable performance loss. Additionally, existing AI-based cell-level mobility management methods focus solely on the prediction accuracy of the reference signal received power (RSRP), ignoring the communication quality degradation resulting from the reactive nature of the event A3-based handover mechanism. Recently, the 3GPP 5G-Advanced New Radio (NR) standardization has launched studies on AI-based mobility management for highly reliable handovers in Release 19.\footnote{\url{https://www.3gpp.org/ftp/tsg_ran/TSG_RAN/TSGR_102/Docs/RP-234055.zip}} Mobility management methods for HSR communications are expected to leverage case-specific characteristics in the design phase or, alternatively, learn from data to yield improved performance in a data-driven manner. 

In this article, we provide two AI-based beam-level and cell-level mobility management approaches for HSR communications. Particularly, by leveraging the sparsity and time-domain correlation of channels in HSR communications, we propose compressed spatial multi-beam measurements via compressive sensing (CS) to improve the spatial-temporal beam prediction accuracy in HSR communications. We also investigate AI-based cell-level mobility management for HSR scenarios and propose a novel AI-assisted cell handover mechanism, performing handover proactively before the signal quality deteriorates by AI-based handover event prediction. A summary of our overview of AI-based mobility enhancement schemes is given in Table \ref{Summary of AI-based mobility management}.

\section{AI-Based Compressed Multi-Beam Measurement for HSR Communications}
\subsection{AI-Based Beam-Level Mobility Management in HSR Communications}
\begin{figure*}[!t]
\centering
\includegraphics[width=7in]{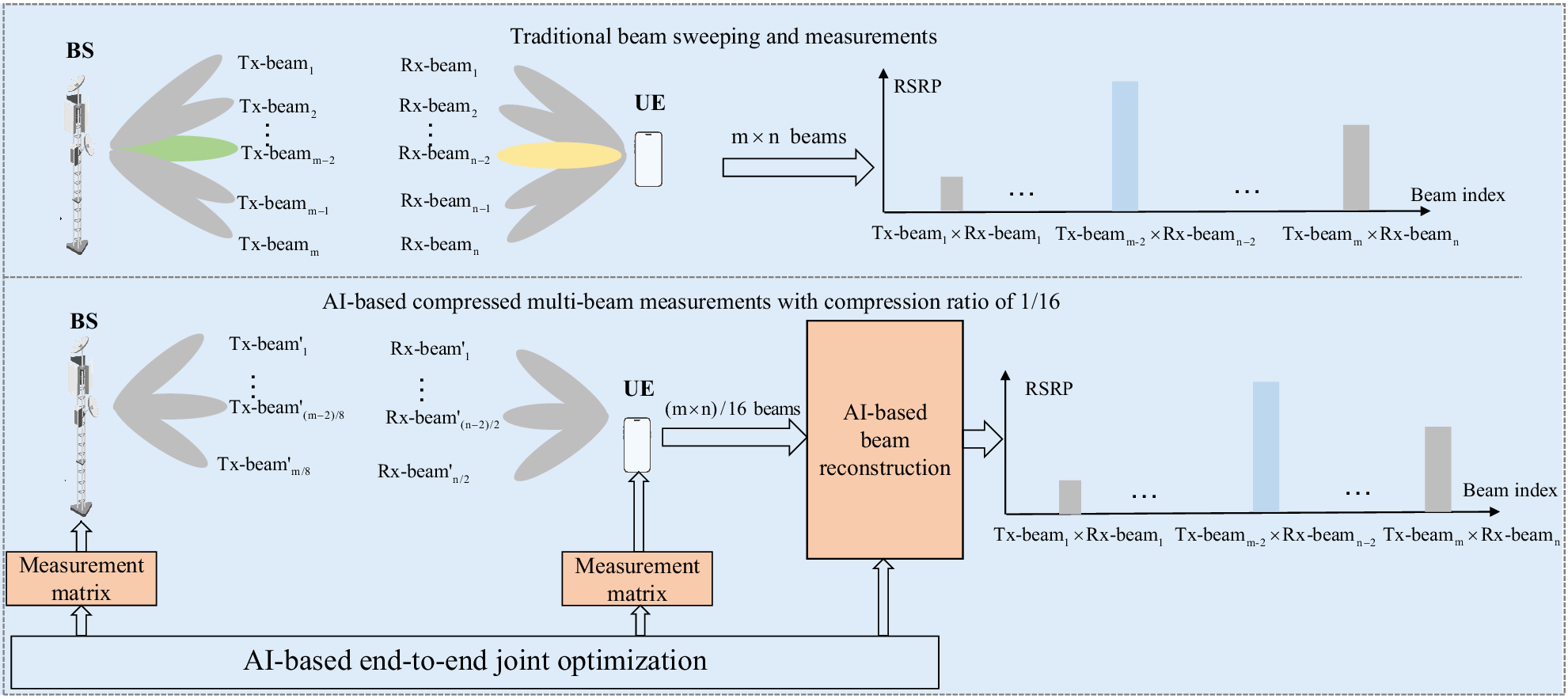}
\caption{Comparison of traditional beam sweeping and beam measurements with AI-based compressed multi-beam measurements.}
\label{Beam management process}
\end{figure*}

When a user equipment (UE) performs a beam switch within the same cell, it initiates beam-level mobility. This process, which includes beam sweeping, beam measurement, beam reporting, and beam determination is managed within the cell, where the physical and media access control layers are sufficient to handle it, thereby avoiding the need for radio resource control (RRC) operations. The detailed beam-level mobility management process is shown in Fig. \ref{Beam management process}. The receiver sequentially scans all predefined beam pairs and evaluates their quality. The beam pair with the highest RSRP is then chosen for beam switching.

Beam-level mobility management enhancement involves three types of beam search methods: traditional methods including exhaustive search and hierarchical search, advanced numerical methods, and AI-based methods. The exhaustive beam search method examines all the candidate beams in the predefined codebook one by one to obtain the beam quality information. The hierarchical beam search method provides a more efficient alternative, which first searches within a wide beam range and then further searches for a narrow beam within the angular range of the selected wide beam. However, traditional methods suffer from high measurement overhead, making them unsuitable for HSR communications, which require frequent beam switching. Numerical methods have also been proposed to address this challenge. For example, \cite{zhang2017codebook} employed tree search for efficient beam training based on a hierarchical codebook, while \cite{seo2015training} proposed an optimal beam training model using Markov decision processes. These numerical approaches mainly target general communication environments, making it challenging to tailor them for particular communication scenarios. In HSR communications, with some coarse-grained beam quality measurements of the serving cell, AI can be utilized to estimate the finely-grained optimal beam pair in the current instance or predict the optimal beam pair in future instances, which leads to reduced measurement overhead and enhanced throughput.

There are two primary use cases for AI models applied in beam-level mobility management, i.e., temporal beam prediction and spatial beam prediction. For temporal beam prediction, where the beam quality information of previous instances is utilized to predict the quality of beams in one or more future instances, most AI-based methods employ recurrent neural networks (RNNs) and their variants, such as LSTM and gated recurrent unit (GRU) models. This is driven by the ability of RNN models to identify temporal patterns in time series data, allowing them to gather extensive information. A two-layer convolutional LSTM network can be used for predicting the beam quality across spatial and temporal dimensions as detailed in \cite{echigo2021deep}. The combination of a CNN and LSTM network was proposed in \cite{10334007} for predicting the optimal beam ID across temporal dimensions. A multi-layer LSTM network was applied in \cite{lim2021deep} for forecasting the optimal future beam. Spatial beam prediction exploits spatial correlations among adjacent beams to estimate the quality of all beams using incomplete beam data. CNNs and networks with fully connected (FC) layers have been commonly employed for this purpose \cite{chen2023active, ma2020machine, alrabeiah2020deep,klautau2019lidar}. Due to the high mobility in HSR communications, the signaling overhead and the time required for frequent beam sweeping become prohibitive. For these reasons, temporal or spatial-temporal beam prediction offers greater benefits in terms of resource conservation and stable performance of HSR communications.

Different tasks in beam-level mobility management require distinct inputs to the AI models in HSR communications. In order to instigate further discussion, there are two vital concepts for beam-level mobility management, namely Set A and Set B.\footnote{\url{https://www.3gpp.org/ftp/tsg\_ran/WG1\_RL1/TSGR1\_109-e/Inbox/drafts/9.2.3.2/R1-220xxxx\%20\%5B109-e-R18-AIML-06\%5D-Summary-1-v022-MediaTek\_Intel.docx}} Set A comprises all narrow beams from the predefined codebook, while Set B encompasses the beams that have been measured and are utilized as inputs into the AI models. Various alternative model input methods exist with respect to the relationship between Set A and Set B. For example, the AI model could receive the input of RSRP measurements of partial narrow beams, making Set B a subset of Set A. This is the most commonly used method in HSR communications, which greatly reduces the measurement overhead. Another option is to utilize RSRP measurements of wide beams to estimate the quality of narrow beams, where Set B and Set A are distinct. In addition, whenever the RSRP measurements of all narrow beams are the inputs into the AI models, Set B can be the same as Set A. The first two types of inputs, where the inputs to the model are the RSRP measurements of some narrow beams and all wide beams, are designed for temporal and spatial beam prediction models, aiming to reduce the measurement overhead by lowering the frequency of beam measurements, which makes them more suitable for latency-sensitive applications. In contrast, the last type of input relies on measuring the RSRP of all candidate beams to conduct temporal beam prediction, achieving the higher model prediction accuracy. This method is well-suited for applications that require exceptionally high accuracy in prediction models and are less sensitive to latency. Instead of using RSRP, supplementary auxiliary information, like the position of the UE, its mobility vector, the signal-to-noise ratio (SNR), and the beam index can currently be utilized to provide broader spatial context to forecast the state of the link connection \cite{na2019deep}.

\subsection{AI-Based Compressed Multi-Beam Measurement Scheme}
In HSR communications, every train follows a predetermined path and performs the same routes periodically. Given sufficient training data, AI models are expected to perform exceptionally well in managing beam-level mobility owing to this intrinsic regularity. For instance, the super-resolution-based spatial beam prediction can be implemented using a sub-pixel convolutional layer paired with an up-sampling layer \cite{echigo2021deep}, or a combination of CNN and FC \cite{chen2023active}. 

Nevertheless, these existing beam prediction solutions rely on the premise that the assessed beams are selected from a predefined subset of all the beams, akin to the down-sampling process. One limitation of beam measurement through down-sampling is that it does not capture any information from the beams that are not sampled, leading to inevitable performance loss. Rather than discarding certain beams outright, a potential approach to reduce this performance loss is to use CS that projects the channel into a lower dimensional space. Given some CS-based beam measurements, deep learning (DL) has been exploited for spatial beam prediction \cite{myers2020deep}. 

\begin{figure*}[!t]
\centering
\includegraphics[width=5in]{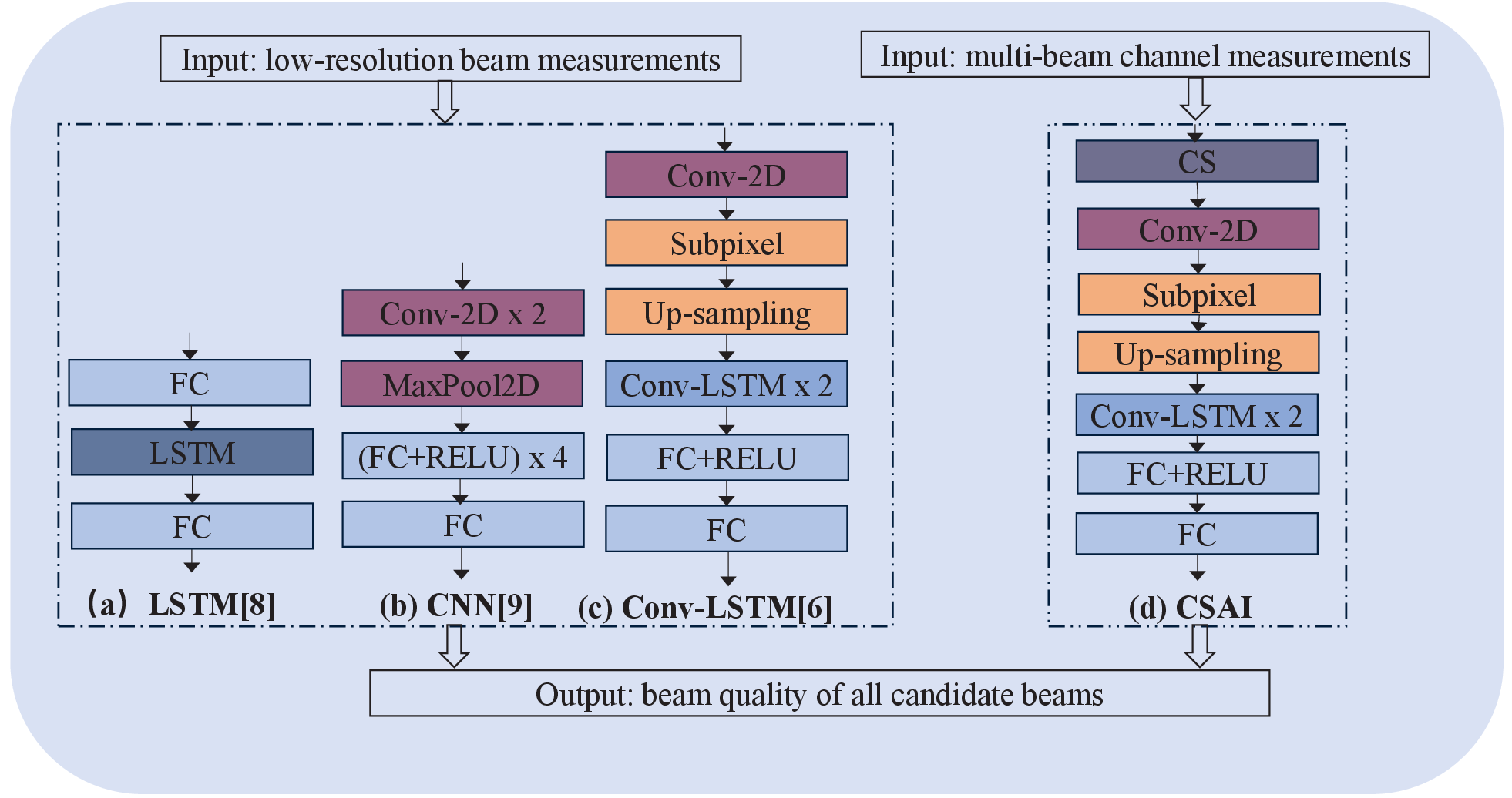}
\caption{AI-based beam-level mobility management methods.} 
\label{Intra_cell_model}
\end{figure*}
Here, we investigate DL-based spatial-temporal beam prediction using compressed multi-beam channel measurements. DL is employed to output the beam quality in the future slots using historical multi-beam channel measurements in the compressed domain, where the CS measurement matrix can be learned jointly. A general model is initially trained and deployed at each cell, where it can undergo periodic retraining or continuous updates using locally collected data. In the training phase, the complete beam channel information from historical instances serves as input into the model. The CS matrix serves as the first layer, namely the linear compression layer. The compressed channel information is fed into the remaining neural network to reconstruct the complete RSRP information of all candidate beams for a future instance. In the remaining module, the compressed features are first processed by a  2D convolution (Conv-2D) layer. This layer extracts spatial features from the compressed representation, ensuring that essential beam information is preserved before further processing. Subsequently, a sub-pixel convolution layer and a 2D upsampling layer are applied before data is fed into the convolutional LSTM (Conv-LSTM) layers. These two layers learn to upsample low-resolution features from limited beam measurements, mapping them to high-resolution feature maps of the complete beam measurements. Then, the initial estimates of the complete beam measurement map are fed into the Conv-LSTM layers to calculate the temporal variations by leveraging historical data and local correlations. Finally, the network outputs a complete RSRP map for an upcoming time slot. As a learnable parameter, the CS measurement matrix is jointly optimized during the training of the model. The structure of the proposed AI model for beam-level mobility management, namely CSAI, is illustrated in Fig. \ref{Intra_cell_model}(d). In the inference phase, the trained measurement matrices are applied at the transceivers while the compressed channel measurements are fed into the trained neural network to conduct beam prediction.
\begin{figure}[t]
\centering
\includegraphics[width=3.2in]{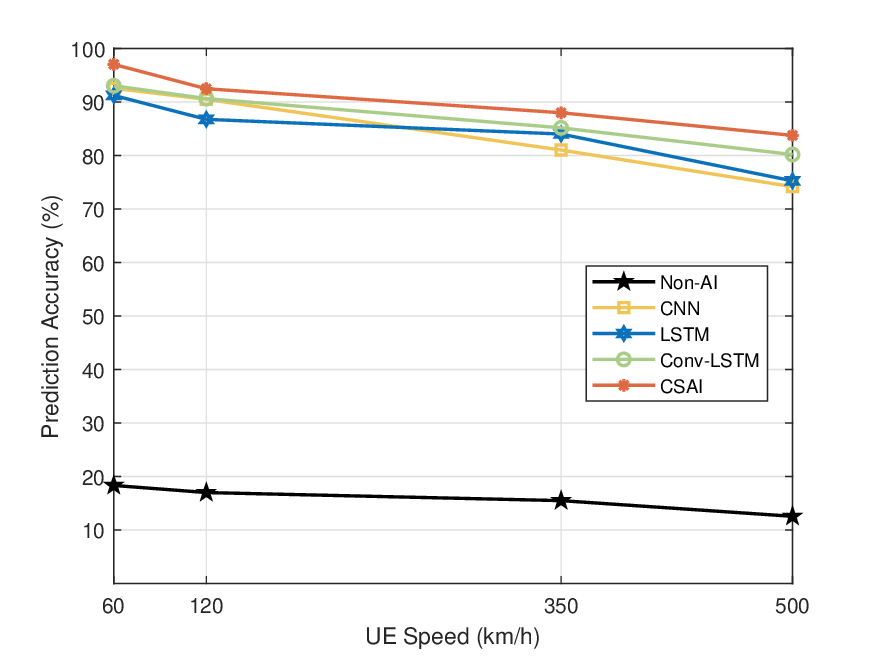}
\caption{Comparison of AI and non-AI methods for spatial-temporal beam prediction in HSR communications.}
\label{intra_cell model performance}
\end{figure}

Now, we compare the performance of AI-based spatial-temporal beam prediction methods, whose structures are shown in Fig. \ref{Intra_cell_model}.  Figure \ref{Intra_cell_model}(a) \cite{lim2021deep} and Fig. \ref{Intra_cell_model}(b) \cite{chen2023active} employ the traditional LSTM for temporal modeling and CNN for spatial feature extraction, respectively, to predict the high-resolution beams. The CSAI model in Fig. \ref{Intra_cell_model}(d) and the AI-based method in Fig. \ref{Intra_cell_model}(c) \cite{echigo2021deep} share the same Conv-LSTM architecture which integrates convolutional operations into the LSTM framework to jointly capture the spatial and temporal features. The only distinction between the two lies in the design of the beam measurement module. In the experiments, the channel is generated using a system-level simulation platform which we build based on QuaDRiGa. We deploy 7 BSs, where each BS is equipped with three half-wavelength spaced uniform plane arrays (UPAs) with a sector orientation of 30, 150 and -90 degrees. We further consider the urban macro scenario for outdoor 30 GHz mmWave communications, while the UE follows a straight-line trajectory which aligns with the HSR scenario. Four distinct datasets, representing UE speeds of 60km/h, 120km/h, 350km/h and 500km/h, are utilized in the evaluation. These datasets comprise 40,000 training samples, 5,000 validation samples, and 5,000 test samples. The batch size is set to 256, and the training runs for a maximum of 800 epochs. The learning rate changes dynamically from $10^{-3}$ to $10^{-8}$, utilizing the adaptive moment estimation optimizer. We consider the prediction accuracy for the best beam with a compression ratio\footnote{The compression ratio is defined as the ratio of the number of input beam measurements to the number of all beam measurements.} of $\frac{1}{16}$. The simple non-AI approach straightforwardly uses the best one out of the partially measured present beams as the prediction. For the three traditional AI-based beam prediction methods shown in Fig. \ref{Intra_cell_model}(a) (b) (c), the model's input uses equal interval sampling from predefined codebook to obtain beam  measurement information with a compression ratio of $\frac{1}{16}$. As shown in Fig. \ref{intra_cell model performance}, the AI-based spatial-temporal beam prediction methods outperform significantly the traditional non-AI method. As the speed of the movement increases, all methods generally show an anticipated degradation in the prediction accuracy. The CSAI demonstrates superior performance for all the cases. At the speed of 500km/h, the CSAI is still able to yield a prediction accuracy of 83.76\%. The network in Fig. \ref{Intra_cell_model}(a) requires 0.145M flops and 0.048M parameters, while the network in Fig. \ref{Intra_cell_model}(b) has 0.146M flops and 0.104M parameters. The Conv-LSTM network in Fig. \ref{Intra_cell_model}(c) demands 3.84M flops and 0.53M parameters. In comparison, the proposed CSAI model uses 3.88M flops and 0.136M parameters, showing that it achieves superior prediction performance with only a minimal increase in computational cost compared to the Conv-LSTM network, which uses a fixed down-sampling beam pattern approach.

\section{AI-Based Proactive Handover Mechanism for HSR Communications}
\subsection{AI-Based Cell-Level Mobility Management in HSR Communications}
Cell-level mobility management deals primarily with cell handovers, initiating the transition of a user's session to a new cell when the signal strength from the serving cell falls below a predefined threshold. Given the rapid movement in HSR communications, cell handovers occur frequently, and failures in these handovers would unavoidably compromise the user experience. Handover decision is performed by the source cell based on UE measurements. UE measurements should meet by certain criteria, for example, event A3 which will be triggered when the difference between the layer 3 RSRP (L3-RSRP)\footnote{L3-RSRP represents the cell-level RSRP, obtained by averaging or maximizing the RSRP of all beam pairs in each cell.} measurement of the target cell and the L3-RSRP measurement of the source cell exceeds a predefined offset. If the above criteria are met during the time to trigger (TTT) duration, the UE reports the measurements and the source cell initiates the handover preparation phase. Afterwards, the cell handover procedure initiates the execution after the UE receives the handover command message, whose task is to provide RRC reconfiguration message to the target cell. The specific cell handover process is shown in Fig. \ref{Handover process}. 

During state 1, the UE measures beam quality and reports it to the source BS while remaining connected to the source cell. In HSR communications, high-speed movement leads to a rapid decline in link quality during this stage, often resulting in issues such as handover failure, radio link failure (RLF), ping-pong effects, and throughput degradation. Leveraging its strong environmental awareness and learning capabilities, AI/ML presents a promising solution to mitigate these challenges by predicting the RSRP during this state.

One approach for AI-based cell-level mobility management is to exploit the beam-level outputs of different cells. In particular, AI models are deployed at the network side, and each cell employs a unique AI model to predict the future RSRP of this cell. The shortcomings of this cell-level mobility management approach lie in the high deployment and management overhead of AI models. Another approach which is more suitable for HSR scenarios is to use an AI model to predict the future RSRP of multiple adjacent cells or the future optimal BS index based on historical RSRP measurement information. In this context, \cite{vankayala2023efficient} utilized a multi-layer LSTM architecture to predict the optimal BS index as the candidate for handover. Moreover, \cite{shah2022multi} developed a CNN-based auto-encoder to reduce the data dimension, and then a LSTM network was used to conduct link quality prediction for each cell. 

For the HSR scenario, we explore various model inputs and outputs tailored to its specific requirements. One option is using the L3-RSRP of each cell as input, with the output being either the L3-RSRP of each cell or the top $K$ cells. L3-RSRP has lower dimensionality than L1-RSRP.\footnote{L1-RSRP represents the beam-level RSRP, which is the RSRP of each beam pair of each cell.} However, calculating L3-RSRP, which involves averaging over multiple time slots, introduces additional latency, undesirable in HSR communications. Another viable approach is to use L1-RSRP as both the input and output of the AI model. The output L1-RSRP can then be processed to derive the L3-RSRP for cell handovers. This method allows L1-RSRP to be used for both beam-level and cell-level mobility management. However, L1-RSRP contains more information than necessary for handovers, and its complex correlation between input and output increases the learning challenge. Moreover, leveraging L1-RSRP to forecast L3-RSRP via AI can be an advantageous strategy. This method reaps the benefits of the above two approaches, making use of the abundant data in RSRP to enhance the AI model's prediction accuracy, while also streamlining the mapping between the model's input and output.

\begin{table*}[!t]
\renewcommand{\arraystretch}{1.3}
\caption{Summary of AI-based mobility management schemes}
\centering
\begin{tabular}{|>{\raggedright\arraybackslash}m{0.3cm}|>{\raggedright\arraybackslash}m{1.2cm}|>{\raggedright\arraybackslash}m{1.3cm}|>{\raggedright\arraybackslash}m{2.8cm}|>{\raggedright\arraybackslash}m{3cm}|>{\raggedright\arraybackslash}m{4.3cm}|>{\raggedright\arraybackslash}m{2cm}|}
\hline
Ref & Category & Network & Input & Output & Description & KPIs\\
\hline
\cite{alrabeiah2020deep} &\multirow{9}{*}[-15ex]{\centering Beam-level}  & FC & The sub-6G channel vector & The optimal beam and the path state like LoS & Utilizing the low-frequency sub-6G channel to predict. & Top-1 accuracy, spectral efficiency\\ \cline{1-1} \cline{3-7}
\cite{na2019deep} &  & FC & Location information, mobility vectors, SNR and current beam & The link connectivity within the next beam tracking phase & Predicting whether there will be a connection loss before the next beam tracking procedure. & Throughput \\ \cline{1-1} \cline{3-7}
\cite{klautau2019lidar} &  & CNN & LIDAR measurements, the coordinates of the BS and UE & The optimal beam and the path state: line-of-sight (LoS) or non line-of-sight (NLoS) & Generating a histogram based on the model's input to learn the channel features. & Accuracy for LoS detection \\ \cline{1-1} \cline{3-7}
\cite{chen2023active} &  & CNN+FC & RSRP of partial narrow beams or all wide beams & The direction of the transceiver beams & Utilizing supervised and unsupervised learning based on CNN and FC. & Achievable rate\\ \cline{1-1} \cline{3-7}
\cite{ma2020machine} &  & CNN+FC & The RSRP of all beams of each UE & The optimal beam index of each UE & Adopting a CNN  to achieve multi-user synchronous alignment. & Spectral efficiency \\ \cline{1-1} \cline{3-7}
\cite{10123939} & & LSTM & The complete RSRP images & The optimal beam index & The multi-layer LSTM network  and FC are utilized. & RSRP difference,Top-K accuracy\\ \cline{1-1} \cline{3-7}
\cite{lim2021deep} &  & LSTM & Channel state information (CSI) and the context vector & CSI and beam quality & The LSTM network and extended Kalman filter (EKF) are used. & Achievable rate\\ \cline{1-1} \cline{3-7}
\cite{10334007} & & CNN+LSTM & The RSRP images of all beam pairs & The optimal beam index of transceiver & Utilizing CNN for feature extraction and LSTM for temporal beam prediction. &RMSE \\ \cline{1-1} \cline{3-7}
\cite{echigo2021deep} &  & Conv-LSTM & The RSRP image of all wide beams & The RSRP image of all narrow beams & Using sub-pixel convolution and Conv-LSTM to realize RSRP prediction. & MSE \\ \cline{1-1} \cline{3-7}
Our &  & CS+AI & Channel information of compressed multi-beam  & RSRP of all beams at the current instance & Learning the optimal measurement matrix and conv-LSTM network for prediction. & Top-1 accuracy\\ 
\hline
\cite{shah2022multi} & \multirow{3}{*}[-5ex]{\centering Cell-level} & CNN+LSTM & SNR of all beams and all cells  & The same as the input & Using CNN to reduce dimension and LSTM network to predict beam quality. &link(s) prediction error\\ \cline{1-1} \cline{3-7}
\cite{vankayala2023efficient} &  & CNN/LSTM & RSRP images of all beams of all cells & The best BS index, and path state (LoS or NLoS) & Using two AI models based on CNN and LSTM predicts the best BS index. & Prediction accuracy of best BS \\ \cline{1-1} \cline{3-7}
Our &  & CNN/LSTM & Compressed RSRP images & RSRP of all cells & Using two AI models to optimize the shortcomings of the handover mechanism. & RLF rates \\ 
\hline
\end{tabular}
\label{Summary of AI-based mobility management}
\end{table*}

\subsection{AI-Based Proactive Handover Scheme}
Existing AI/ML-assisted cell handover methods typically use the RSRP of all beams across cells to predict the future RSRP for each beam or to identify the optimal base station index for upcoming handovers \cite{vankayala2023efficient,shah2022multi}. These methods primarily focus on the improvements in prediction accuracy and spectral efficiency provided by AI models, with limited attention to system-level metrics, such as RLF rates and PP rates. This gap may overlook the potential impact of model design on system reliability and service continuity. Additionally, existing algorithms fail to address the limitations of the event A3-based handover mechanism, which is a passively reactive approach triggered only after the UE detects irreversible communication quality degradation.

\begin{figure*}[t]
\centering
\includegraphics[width=7in]{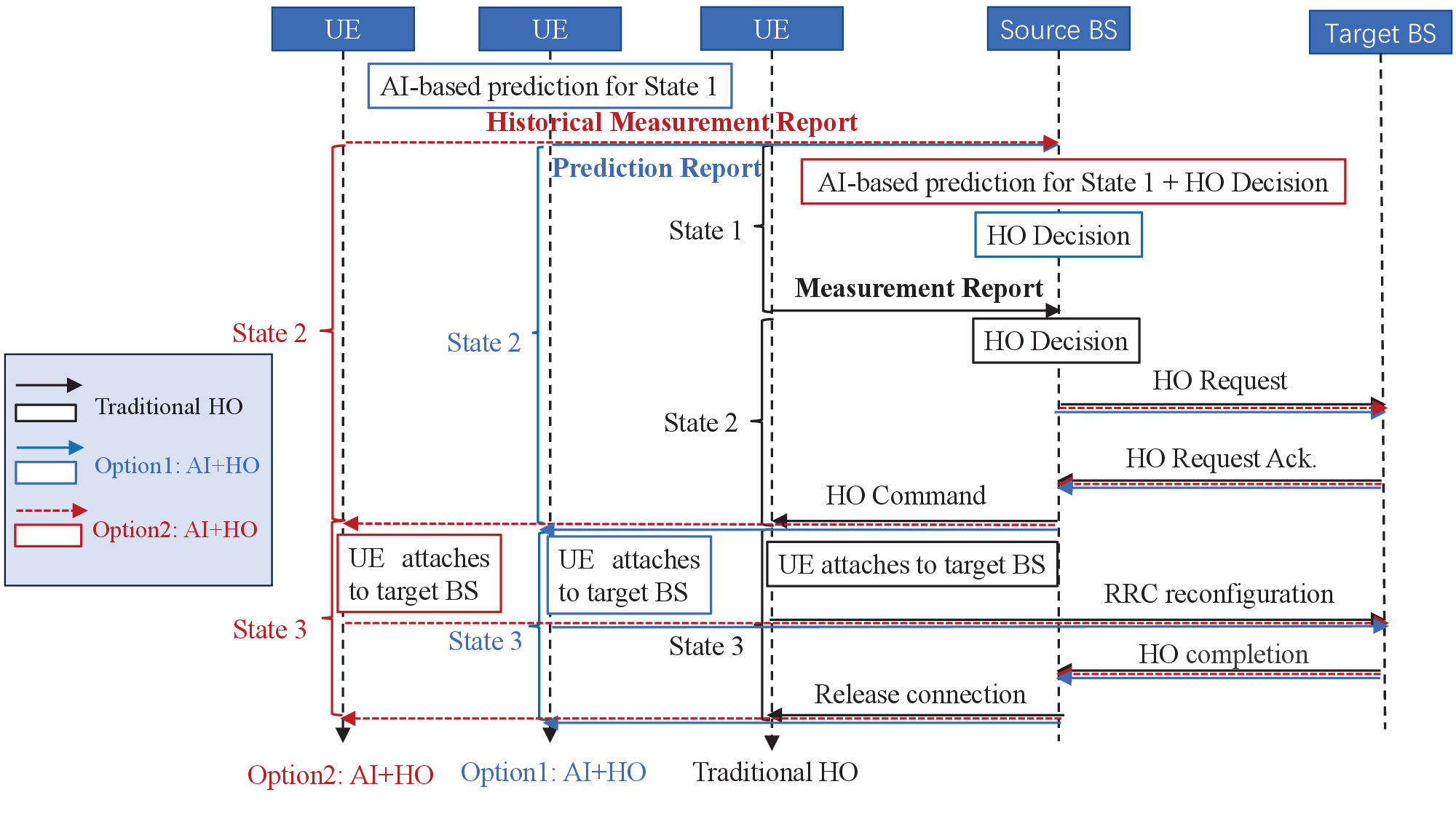}
\caption{Comparison of an AI-assisted cell handover process with traditional cell handover process.}
\label{Handover process}
\end{figure*}

To maintain a stable communication quality for HSR communications, we provide a novel AI-assisted cell handover scheme. We use an AI model to predict the L3-RSRP of all cells for future state 1 based on historical L1-RSRP to aid the source BS in handover decisions. Using this method, we predict the future communication quality variation trends of adjacent cells and proactively switch cells based on these predictions. This approach effectively mitigates the communication quality degradation and reduces handover time from the original states (state 1, state 2, state 3) to the existing states (state 2, state 3). The AI-based cell handover process is illustrated in Fig. \ref{Handover process}, where option 1 and option 2 represent different deployment methods of AI models. In option 1, the AI-based prediction model is deployed on the UE side, with the prediction report uploaded to the source BS to assist in the handover decision. In option 2, the AI-based prediction model is deployed on the network side, where the network performs temporal prediction and makes the handover decision based on historical measurement reports. The primary difference between the two options lies in the model's deployment mode, while the model's inputs, outputs, and architecture remain the same.In HSR scenarios, option 1 is more suitable as it directly uploads the predicted results instead of raw measurement data, reducing the transmission overhead and saving time-frequency resources. 

Based on the proposed handover scheme, we explore two approaches to reduce measurement overhead. One uses all beams from partially selected cells, while the other uses partially selected beams of each cell to predict the L3-RSRP for all cells. The model input is obtained through equal-interval sampling rather than the compressed multi-beam measurement proposed in Section II to independently evaluate the performance gains of the AI-assisted handover scheme. We utilize the partial L1-RSRP from the past six time instances to predict the L3-RSRP in the next four time instances. Meanwhile, we employ CNN-based and LSTM-based models for RSRP prediction. The LSTM model captures temporal patterns in handover-related data and consists of four LSTM layers and two FC layers, with four parallel networks predicting L3-RSRP at different future instances. The CNN model, offering lower complexity and faster inference, includes 2D convolution layers with 12, 24, and 32 kernels, followed by two FC layers. While this study adopts classical AI architectures such as LSTM, CNN, and Conv-LSTM to validate the effectiveness of AI-driven mobility management, these models can be replaced in the future by more advanced architectures—such as large language models—to further enhance the performance of AI-driven beam-level and cell-level mobility management.

The L1-RSRP and L3-RSRP in our experiments are generated using a system-level simulation platform introduced in Section II.B. We have uploaded two datasets to GitHub with detailed descriptions and scenario configurations.\footnote{\url{https://github.com/wc253/dataset_mobility_management}} Beam-level and cell-level setups differ mainly in how they handle interference. Beam-level experiments focus on intra-cell beam switching and disregard inter-cell interference. In contrast, cell-level experiments involve inter-cell handover and account for inter-cell interference by modeling it stochastically in our setup. Each dataset comprises 50,000 samples, which are split into training, validation, and test sets in a 3:1:1 ratio. For the proposed two approaches with equidistant sampling, L1-RSRP of 50\% beams in each cell or all beams in 50\% cells are selected as input to the AI model. The batch size is 256 and the maximum epoch is 800. The learning rate dynamically changes from $10^{-3}$ to $10^{-8}$, utilizing the adaptive moment estimation optimizer. The RSRP difference is one evaluation metric of prediction accuracy for cell-level mobility management, which reflects the disparity between the predicted L3-RSRP and the actual L3-RSRP of all cells. In addition, we also consider the RLF rates as the system-level evaluation metric to illustrate the overall handover performance of the communication system. The experimental results show that both CNN-based and LSTM-based models show excellent cell-level temporal prediction performance. Table \ref{RLF rates of AI and Non-AI} shows the significantly reduced RLF rates in AI-assisted cell handover processes. Remarkably, the proposed two approaches to reduce the measurement overhead for cell-level beam management, labeled as Part\_Cell and Part\_Beam, achieve comparable RLF performance with the traditional approach using all beam measurements of all cells, labeled as All\_Beam\_Cell, while we can save 50\% beam measurement overhead for HSR communications. Furthermore, as shown in Table \ref{RLF rates of AI and Non-AI}, the number of flops and total parameters of the CNN-based model are much lower than those of the LSTM-based model, while achieving comparable performance.

\begin{table*}[!t]
\renewcommand{\arraystretch}{1.3}
\caption{Comparison of the RLF Rates and Complexity of AI-based and Non-AI cell-level mobility management methods}
\centering
\begin{tabular}{c|c|c|c|c|c|c}
\hline
\multirow{2}{*}{} & \multicolumn{4}{c|}{UE Speed} & \multirow{2}{*}{Flops} & \multirow{2}{*}{Total Params} \\
\cline{2-5}
 & 60km/h & 120km/h & 350km/h & 500km/h & & \\
\hline
Non-AI & 24.61\% & 27.75\% & 39.06\% & 39.91\% & $\backslash$ & $\backslash$ \\
\hline
CNN\_ALL\_Beam\_Cell & 5.08\% & 9.44\% & 10.07\% & 10.07\% & 33.85M & 28.66M \\
\hline
CNN\_Part\_Cell & 5.39\% & 9.39\% & 9.84\% & 10.33\% & 5.59M & 3.61M \\
\hline
CNN\_Part\_Beam & 5.36\% & 9.26\% & 10.23\% & 10.45\% & 4.82M & 2.51M \\
\hline
LSTM\_ALL\_Beam\_Cell & 5.08\% & 9.57\% & 10.06\% & 10.22\% & 391.87M & 66.02M \\
\hline
LSTM\_Part\_Cell & 5.32\% & 9.59\% & 10.07\% & 10.43\% & 354.46M & 58.46M \\
\hline
LSTM\_Part\_Beam & 5.08\% & 9.58\% & 9.87\% & 10.22\% & 348.52M & 58.80M \\
\hline
\end{tabular}
\label{RLF rates of AI and Non-AI}
\end{table*}

\section{Opportunities and Challenges of AI-Based Mobility Management for HSR Communications}

\subsection{Dynamic Interference Management}
Due to the high mobility in HSR communications, the randomness of inter-cell interference constitutes a crucial challenge. Specifically, mmWave beams from neighboring cells directed at local terminals can cause severe inter-cell interference, while frequent cell handovers and beam switching exacerbate the randomness of this interference. Thus, agile interference management is essential for reliable service. As a potential solution, graph neural networks (GNNs) can intrinsically match the topology of wireless networks and can be used to effectively handle dynamic interference management.

\subsection{Multi-Criteria Joint Decision-Making Mechanism for Mobility Management}
In HSR environments, rapid movement demands adaptive mobility management strategies to handle frequent environmental changes. Relying solely on RSRP variations is inadequate for stable handover decisions. A multi-criterion decision-making mechanism is essential, incorporating factors such as signal interference, network load, terminal velocity, and movement direction alongside RSRP. Integrating these inputs into AI models enhances handover accuracy and reliability, ensuring seamless communication for high-speed trains. Future research could further improve the multi-criteria joint decision mechanism by exploring advanced optimization techniques, such as reinforcement learning-based adaptive weighting of criteria, or leveraging federated learning to enhance model generalization across different HSR environments. 

\subsection{Model Deployment and Compression}
AI-assisted mobility management in practical scenarios demands high energy consumption due to its computational complexity, making network-side deployment more viable. In this setup, the UE reports measurement data to the source BS, which processes it using AI/ML models for mobility management. However, the transmission introduces additional latency, and inevitable reporting errors can degrade the AI model`s performance. Deploying AI models on the UE side could mitigate these issues. Leveraging advanced AI models, such as lightweight architectures or transformer-based designs, can enhance computational efficiency while reducing UE hardware requirements.

\subsection{Model Generalization Design}
AI/ML models designed for specific scenarios would experience performance degradation when applied to different railroad environments due to variations in factors such as BS height, channel characteristics, and railway layouts. To mitigate this issue, the most effective approach is to collect new datasets from the target environment and either retrain the model or apply transfer learning to fine-tune it. This enables the model to better adapt to the new conditions while minimizing performance loss. Furthermore, continuous learning is crucial in dynamic environments, allowing models to adapt to changing channel conditions and movement patterns in real time. However, improving model generalization through continuous learning introduces significant deployment overhead. Balancing these factors remains a key challenge in HSR communication.
 
\subsection{Model Explainability}
Explainability is essential to ensure model trustworthiness, safety, and effectiveness. However, achieving explainability in HSR communications presents several challenges. First, HSR communications demand high accuracy in decision-making. More complex models usually provide better performance but tend to be more opaque and difficult to interpret. Striking a balance between performance and explainability remains a key challenge. Moreover, complex interpretability methods can increase computational overhead, potentially affecting the communication system's response time. Therefore, finding a balance between explainability and real-time performance is a critical research direction in HSR communications.

\section{Conclusion}
This article explored AI-based beam-level and cell-level mobility management in HSR communications. Firstly, a hybrid approach, combining AI and CS, demonstrated superior performance compared to conventional AI-based methods with the same measurement overhead. Secondly, we unveiled the significant gain of AI-based methods for proactive handover, while the measurement overhead can be reduced by using partial beam measurements, without compromise of the RLF rates. Lastly, the opportunities and challenges on AI-based mobility management were briefly discussed for further studies.

\bibliographystyle{IEEEtran}
\bibliography{IEEEabrv,ref}

\begin{thebibliography}{10}
\providecommand{\url}[1]{#1}
\csname url@samestyle\endcsname
\providecommand{\newblock}{\relax}
\providecommand{\bibinfo}[2]{#2}
\providecommand{\BIBentrySTDinterwordspacing}{\spaceskip=0pt\relax}
\providecommand{\BIBentryALTinterwordstretchfactor}{4}
\providecommand{\BIBentryALTinterwordspacing}{\spaceskip=\fontdimen2\font plus
\BIBentryALTinterwordstretchfactor\fontdimen3\font minus \fontdimen4\font\relax}
\providecommand{\BIBforeignlanguage}[2]{{%
\expandafter\ifx\csname l@#1\endcsname\relax
\typeout{** WARNING: IEEEtran.bst: No hyphenation pattern has been}%
\typeout{** loaded for the language `#1'. Using the pattern for}%
\typeout{** the default language instead.}%
\else
\language=\csname l@#1\endcsname
\fi
#2}}
\providecommand{\BIBdecl}{\relax}
\BIBdecl

\bibitem{3gpp2020}
{3GPP}, ``{{NR} and {NG-RAN} {overall description}; Stage 2, v.17.8.0},'' {3GPP}, Tech. Rep. TS 38.300, Mar. 2024.

\bibitem{10123939}
\emph{\emph{Q. Li}} \emph{et al.}, ``{Machine learning based time domain millimeter-wave beam prediction for 5G-advanced and beyond: Design, analysis, and over-the-air Experiments},'' \emph{IEEE J. Sel. Areas Commun.}, vol.~41, no.~6, pp. 1787--1809, Jun. 2023.

\bibitem{shah2022multi}
S.~H.~A. Shah and S.~Rangan, ``{Multi-cell multi-beam prediction using auto-encoder LSTM for mmWave systems},'' \emph{IEEE Trans. Wireless Commun.}, vol.~21, no.~12, pp. 10\,366--10\,380, Dec. 2022.

\bibitem{zhang2017codebook}
J.~Zhang, Y.~Huang, Q.~Shi, J.~Wang, and L.~Yang, ``{Codebook design for beam alignment in millimeter wave communication systems},'' \emph{IEEE Trans. Commun.}, vol.~65, no.~11, pp. 4980--4995, Nov. 2017.

\bibitem{seo2015training}
J.~Seo, Y.~Sung, G.~Lee, and D.~Kim, ``{Training beam sequence design for millimeter-wave MIMO systems: A POMDP framework},'' \emph{IEEE Trans. Signal Process.}, vol.~64, no.~5, pp. 1228--1242, Mar. 2016.

\bibitem{echigo2021deep}
H.~Echigo, Y.~Cao, M.~Bouazizi, and T.~Ohtsuki, ``{A deep learning-based low overhead beam selection in mmWave communication},'' \emph{IEEE Trans. Veh. Technol.}, vol.~70, no.~1, pp. 682--691, Jan. 2021.

\bibitem{10334007}
Y.~Bai, J.~Zhang, C.~Sun, L.~Zhao, H.~Li, and X.~Wang, ``{AI-based beam management in 3GPP: Optimizing data collection time window for temporal beam prediction},'' \emph{IEEE Open J. Veh. Technol.}, vol.~5, pp. 48--55, Dec. 2023.

\bibitem{lim2021deep}
S.~H. Lim, S.~Kim, B.~Shim, and J.~W. Choi, ``{Deep learning-based beam tracking for millimeter-wave communications under mobility},'' \emph{IEEE Trans. Commun.}, vol.~69, no.~11, pp. 7458--7469, Nov. 2021.

\bibitem{chen2023active}
H.~Chen, L.~Liu, S.~Xue, Y.~Sun, and J.~Pang, ``{Active sensing for beam management: A deep-learning approach},'' in \emph{Proc. IEEE WCNC}, Mar. 2023, pp. 1--6.

\bibitem{ma2020machine}
W.~Ma, C.~Qi, and G.~Y. Li, ``{Machine learning for beam alignment in millimeter wave massive MIMO},'' \emph{IEEE Wireless Commun. Lett.}, vol.~9, no.~6, pp. 875--878, Jun. 2020.

\bibitem{alrabeiah2020deep}
M.~Alrabeiah and A.~Alkhateeb, ``{Deep learning for mmWave beam and blockage prediction using sub-6 GHz channels},'' \emph{IEEE Trans. Commun.}, vol.~68, no.~9, pp. 5504--5518, Sept. 2020.

\bibitem{klautau2019lidar}
A.~Klautau, N.~González-Prelcic, and R.~W. Heath, Jr., ``{LIDAR data for deep learning-based mmWave beam-selection},'' \emph{IEEE Wireless Commun. Lett.}, vol.~8, no.~3, pp. 909--912, Jun. 2019.

\bibitem{na2019deep}
W.~Na, B.~Bae, S.~Cho, and N.~Kim, ``{Deep-learning based adaptive beam management technique for mobile high-speed 5G mmWave networks},'' in \emph{Proc. IEEE ICCE}, Sept. 2019, pp. 149--151.

\bibitem{myers2020deep}
N.~J. Myers, Y.~Wang, N.~González-Prelcic, and R.~W. Heath, Jr., ``{Deep learning-based beam alignment in mmwave vehicular networks},'' in \emph{Proc. IEEE ICASSP}, May 2020, pp. 8569--8573.

\bibitem{vankayala2023efficient}
\emph{\emph{S. Vankayala}} \emph{et al.}, ``{Efficient deep-learning models for future blockage and beam prediction for mmWave systems},'' in \emph{Proc. IEEE/IFIP NOMS}, May 2023, pp. 1--8.

\end{thebibliography}

\newpage

\vfill

\end{document}